\documentclass[12pt]{article}
\usepackage{latexsym}
\usepackage{graphics}

\begin{document}


\title{Scalar-tensor  black holes  coupled to Euler-Heisenberg nonlinear
electrodynamics}

\author{Ivan Zh. Stefanov\thanks{E-mail: zhivkov@phys.uni-sofia.bg}, \,\,
     Stoytcho S. Yazadjiev \thanks{E-mail: yazad@phys.uni-sofia.bg}\\
{\footnotesize  Dept. of Theoretical Physics,
                Faculty of Physics}\\ {\footnotesize St.Kliment Ohridski University of Sofia}\\
{\footnotesize  5, James Bourchier Blvd., 1164 Sofia, Bulgaria }\\\\[-3.mm]
 Michail D.~Todorov\thanks{E-mail: mtod@tu-sofia.bg}
\\ [-1.mm]{\footnotesize
{Faculty of Applied Mathematics and Computer Science}}\\
[-1.mm] {\footnotesize {Technical University of Sofia}}\\
[-1.mm] {\footnotesize 8, Kliment Ohridski Blvd., 1000 Sofia, Bulgaria}}

\date{}

\maketitle

\begin{abstract}
The no-scalar-hair conjecture rules out the existence of asymptotically flat black holes with a scalar dressing for
a large class of theories. No-scalar-hair theorems have been proved for the cases of neutral black holes and for
charged black holes in the Maxwell electrodynamics. These theorems, however,
do not apply in the case of non-linear electrodynamics.
In the present work numerical solutions describing
charged black holes coupled to Euler-Heisenberg type non-linear electrodynamics in scalar-tensor theories of gravity
with massless scalar field are found. In comparison to the corresponding solution in General Relativity the
presented solution has a simpler causal structure the reason for which is the presence of the scalar field.
The present class of black holes has a single, non-degenerate horizon, i.e.,
its causal structure resembles that of the Schwarzschild black hole.

Pacs numbers: 4.25.Dm, 04.40.-b, 04.70.-s, 95.30.Tg, 97.60.Lf
\end{abstract}




\sloppy

\section{Introduction}

Scalar-tensor theories of gravity are the most natural generalization of General  Relativity (GR)
and arise naturally in string theory and higher dimensional gravity theories \cite{DP}. Different modifications of
scalar-tensor theories are attracting much interest also in cosmology and astrophysics.
The search of black holes in the scalar-tensor theories leads to the origin of the no-scalar-hair conjecture.
It is similar to the no-hair conjecture in GR which states that in the exterior of a black hole
the only information available regarding the black hole may be that of its mass, charge, and angular momentum.
For neutral static, spherically symmetric black holes the only available information is about its mass, and the
exterior of the black hole is reduced to the Schwarzschild solution.
According to the above mentioned no-scalar-hair conjecture the presence of a scalar field would not lead
to the existence of other preserved
quantities which would allow a distant observer to distinguish between the Schwarzschild black hole
and a black hole with a scalar dressing.

A no-scalar-hair theorem which rules out the existence of static, spherically symmetric,
asymptotically flat, neutral black holes with regular, non-trivial scalar field was proved by Saa \cite{Saa}.
This theorem holds for a large class of
scalar tensor theories in which the scalar field is non-minimally coupled to gravity.
In order to prove the theorem Saa applied an explicit, covariant method to generate the exterior solutions for
these theories through conformal mapping of solutions from the minimally coupled case. The
scalar field in these theories becomes a constant and hence trivial if one demands that the essential
singularity at the center of symmetry is hidden by an event horizon. A similar theorem treating also the
case of charged scalar field with self-interaction was proved by Bekenstein \cite{Be}.

Banerjee and Sen \cite{BSen} generalized Saa's theorem for the case of charged black holes in Maxwell electrodynamics.

In the case of non-linear electrodynamics, however, the energy-momentum tensor of the electromagnetic field
has a non-zero trace a sequence of which is that the electromagnetic field is non-trivially coupled to the scalar field. Hence,
we can expect that the no-scalar-hair theorems might not hold in that case. In the present work we prove that
our assumption is correct
and find numerical solutions describing black holes with a non-trivial scalar field coupled to non-linear electrodynamics.

We consider a particular example of nonlinear electrodynamics -- the Euler-Heisenberg electrodynamics.
The nonlinear electrodynamics was first introduced by Born and Infeld in 1934 to obtain finite energy density
model for the electron \cite{BI}. In recent years nonlinear electrodynamics models are attracting much interest, too.
The reason is that the nonlinear electrodynamics arises naturally in open strings and $D$-branes \cite{L}.
Nonlinear electrodynamics models coupled to gravity and stringy non-linear electrodynamics
have been discussed  in different aspects
(see, for example, \cite{Demianski}-\cite{Y}
and references therein).Here we obtain and discuss black hole solutions coupled to the nonlinear Euler-Heisenberg electrodynamics within
scalar-tensor theories.

Asymptotically flat black
holes in Euler-Heisenberg theory coupled to Einstein gravity were studied in \cite{YT}.

The effective Lagrangian for electrodynamics due to one-loop quantum corrections was calculated by Heisenberg and Euler
\cite{EH}:

\begin{equation}
L_{\rm EH} = -{1\over 4}F_{\mu\nu}F^{\mu\nu} + \frac{1}{4} \,b^2 \left(F_{\mu\nu}F^{\mu\nu}\right)^2 + \gamma \left[ F_{\mu\nu}(\star F)^{\mu\nu} \right]^2
\end{equation}
where $b^2= {8he^4/ (2880 \pi^2 m^4)}$, $\gamma = {7he^4/(5760\pi^2m^4)}$ and $h$, $e$, and $m$ are the Planck constant,
electron charge, and electron mass, respectively, and the star ``$\star$'' stands for the Hodge operator. The star sign denotes the Hodge dual. From experimental aspect,
the Euler-Heisenberg theory is more accurate classical approximation of QED than the Maxwell theory when the field has
high intensity\cite{SB}. Regarding the electric-magnetic duality, the Euler-Heisenberg action breaks it as it was
pointed in \cite{GR1}.

\section{Formulation of the problem}

The general form of the extended gravitational action in
scalar-tensor theories is

\begin{eqnarray} \label{JFA}
S = {1\over 16\pi G_{*}} \int d^4x \sqrt{-{\tilde
g}}\left[{F(\Phi)\tilde {\cal R}} - Z(\Phi){\tilde
g}^{\mu\nu}\partial_{\mu}\Phi
\partial_{\nu}\Phi  \right. \nonumber  \\ \left. -2 U(\Phi) \right] +
S_{m}\left[\Psi_{m};{\tilde g}_{\mu\nu}\right] .
\end{eqnarray}
Here, $G_{*}$ is the bare gravitational constant, ${\tilde R}$ is
the Ricci scalar curvature with respect to the space-time metric
${\tilde g}_{\mu\nu}$. The dynamics of the scalar field $\Phi$
depends on the functions $F(\Phi)$, $Z(\Phi)$ and $U(\Phi)$. In
order that the gravitons  carry positive energy the function
$F(\Phi)$ must be positive. The nonnegativity of the energy of
the scalar field requires that $2F(\Phi)Z(\Phi) +
3[dF(\Phi)/d\Phi]^2 \ge 0$.
The action of matter should depend only on the
material fields $\Psi_{m}$ and the space-time metric ${\tilde
g}_{\mu\nu}$ but not on the scalar field $\Phi$ so that the weak equivalence principle is satisfied.

For convenience, it is a standard mathematical technique to study the scalar-tensor theory
in the conformally related Einstein frame
 given by the metric:
\begin{equation}\label {CONF1}
g_{\mu\nu} = F(\Phi){\tilde g}_{\mu\nu} . \label{conf}
\end{equation}
The transition to the Einstein conformal frame includes not only a conformal transformation of the metric
but also a redefinition of the scalar field. The new scalar field in the Einstein frame
is defined in the following way:

\begin{equation}\label {CONF2}
\left(d\varphi \over d\Phi \right)^2 = {3\over
4}\left\{{d\ln[F(\Phi)]\over d\Phi } \right\}^2 + {Z(\Phi)\over 2
F(\Phi)}
\end{equation}
and

\begin{equation}\label {CONF3}
{\cal A}(\varphi) = F^{-1/2}(\Phi) \,\,\, ,\nonumber \\
2V(\varphi) = U(\Phi)F^{-2}(\Phi) .
\end{equation}
In the Einstein frame action (\ref{JFA}) takes the form

\begin{eqnarray}\label{EFA}
S= {1\over 16\pi G_{*}}\int d^4x \sqrt{-g} \left[{\cal R} -
2g^{\mu\nu}\partial_{\mu}\varphi \partial_{\nu}\varphi -
4V(\varphi)\right] \nonumber \\ + S_{m}[\Psi_{m}; {\cal
A}^{2}(\varphi)g_{\mu\nu}]
\end{eqnarray}
where $R$ is the Ricci scalar curvature with respect to the
Einstein metric $g_{\mu\nu}$.

We take the following Jordan frame nonlinear electrodynamics action

\begin{equation}
S_{m} = {1\over 4\pi G_{*}}\int d^4x \sqrt{{-\tilde g}} L(X, Y)
\end{equation}
where

\begin{equation}
X = {1\over 4} F_{\mu\nu}{\tilde g}^{\mu\alpha} {\tilde g}^{\nu\beta} F_{\alpha\beta},  \,\,\,
Y = {1\over 4}  F_{\mu\nu}\left({\tilde \star} F\right)^{\mu\nu}
\end{equation}
and ${\tilde \star}$ is the Hodge dual with respect to the Jordan frame metric ${\tilde g}_{\mu\nu}$.

In the Einstein frame we have

\begin{equation}\label{EFNEDA}
S_{m} = {1\over 4\pi G_{*}}\int d^4x \sqrt{-g} {\cal A}^4(\varphi) L(X, Y)
\end{equation}
where

\begin{equation}
X = {{\cal A}^{-4}(\varphi)\over 4} F_{\mu\nu}{g}^{\mu\alpha} {g}^{\nu\beta} F_{\alpha\beta}, \label{X}  \,\,\,
Y = {{\cal A}^{-4}(\varphi)\over 4}  F_{\mu\nu}\left({ \star} F\right)^{\mu\nu}
\end{equation}
and ``$\star$'' is the Hodge dual with respect to the Einstein frame metric $g_{\mu\nu}$.

Through varying the action (\ref{EFA}) with (\ref{EFNEDA}) we obtain the following field equations

\begin{eqnarray}
&&{\cal R}_{\mu\nu} = 2\partial_{\mu}\varphi \partial_{\nu}\varphi +  2V(\varphi)g_{\mu\nu} -
 2\partial_{X} L(X, Y) \left(F_{\mu\beta}F_{\nu}^{\beta} -
{1\over 2}g_{\mu\nu}F_{\alpha\beta}F^{\alpha\beta} \right)  \nonumber \\
&&-2{\cal A}^{4}(\varphi)\left[L(X,Y) -  Y\partial_{Y}L(X, Y) \right] g_{\mu\nu}, \nonumber  \\
&&\nabla_{\mu} \left[\partial_{X}L(X, Y) F^{\mu\nu} + \partial_{Y}L(X, Y) (\star F)^{\mu\nu} \right] = 0 \label{F},\\
&&\nabla_{\mu}\nabla^{\mu} \varphi = {dV(\varphi)\over d\varphi } -
4\alpha(\varphi){\cal A}^{4}(\varphi) \left[L(X,Y) -  X\partial_{X}L(X,Y) -  Y\partial_{Y}L(X, Y) \right], \nonumber
\end{eqnarray}
where $\alpha(\varphi) = {d\ln{\cal A}(\varphi)\over d\varphi}$.

In what follows we consider the truncated\footnote{Here we consider the pure magnetic and the pure
electric  case for which $Y=0$.  }
Euler-Heisenberg  electrodynamics described by the Lagrangian

\begin{equation}
L_{\rm EH}(X) = -  X + 4b^2 X^2 . \label{EHL}
\end{equation}

\section{Basic equations and qualitative investigation}

In this paper, we will be searching for black hole solutions, namely solutions
which have an event horizon on which the scalar field $\varphi$ is regular.
In order to ensure the smooth transition between the Einstein and the Jordan  conformal frames, we will impose some
restricting conditions on the coupling function ${\cal A}(\varphi)$. We will require that $0<{\cal A}(\varphi)<\infty$
for $r_{H}\leq r \leq \infty$, where $r_{H}$ is the radius of the horizon.
The coupling function ${\cal A}(\varphi)$ (respectively $\alpha(\varphi)$) determines the properties of the
solutions and contains the diversity of scalar-tensor theories.
In the present work we will consider only theories for which $\alpha(\varphi)$ has a fixed positive sign for all values
of $\varphi$. The manner of investigation of solutions within theories with negative $\alpha(\varphi)$ is similar.
Theories in which
the coupling function changes its sign are much more complicated (also from numerical point of view)
since in them some interesting effects like
bifurcation of solutions can appear, especially when $\alpha(\varphi)\sim\varphi$. Such solutions
are currently being studied by the authors and the results will be given elsewhere.

\subsection{Magnetically charged black holes}

The metric of a static, spherically symmetric spacetime can be written in the form

\begin{equation}
ds^2 = g_{\mu\nu}dx^{\mu}dx^{\nu} = - f(r)e^{-2\delta(r)}dt^2 + {dr^2\over f(r) } +
r^2\left(d\theta^2 + \sin^2\theta d\phi^2 \right).
\end{equation}
In the magnetically charged case the electromagnetic field is given by

\begin{equation}
F = P \sin\theta d\theta \wedge d\phi
\end{equation}
and the magnetic charge is denoted by $P$.

The field equations reduce to the following coupled system of
ordinary differential equations:
\begin{eqnarray}
&&\frac{d\delta}{dr}=-r\left(\frac{d\varphi}{dr} \right)^2\label{MagnDelta},\\
&&\frac{d m}{dr}=r^2\left[\frac{1}{2}f\left(\frac{d\varphi}{dr}\right)^2-{\cal A (\varphi)}^{4}L(X) \right],\\ \label{MagnM}
&&\frac{d }{dr}\left( r^{2}f\frac{d\varphi }{dr} \right)=
r^{2}\left\{-4\alpha(\varphi){\cal A}^{4}(\varphi) \left[L(X) -  X\partial_{X}L(X)\right] -
r f\left(\frac{d\varphi}{dr} \right)^3    \right\} \label{MagnPhi}  ,
\end{eqnarray}
where $ X $ reduces to:
\begin{equation}
X = {{\cal A}^{-4}(\varphi)\over 2} \frac{P^2}{r^4}.
\end{equation}

\subsection{Electrically charged black holes}

In the electrically charged case we make the following ansatz for the electromagnetic field
\begin{equation}
F = F_{\rm tr} dt \wedge dr,
\end{equation}
and through eq.(\ref{X}) and the equation for the electromagnetic field in (\ref{F})
we obtain the following equation for $X$,
\begin{equation}
64b^4 X^{3}-16b^2X^2+X=-\frac{1}{2}\frac{Q^2}{r^4{\cal A}^{4}(\varphi)}.
\end{equation}
It has one real root, namely
\begin{equation}
X=\frac{1}{24 b^2}\left( h^{1/3}+h^{-1/3}+2 \right),\label{Xsol}
\end{equation}
where
\begin{equation}
h=-1-\frac{54 Q^2 b^2}{r^4{\cal A}^{4}(\varphi)} + 6b\sqrt{3}\sqrt{\frac{Q^2\left[1+\frac {27 Q^2 b^2}
{r^4{\cal A}^{4}(\varphi)}\right]}{r^4{\cal A}^{4}(\varphi)}}.
\end{equation}
In the pure electric case eqs.(\ref{F}) reduce to the
following system of non-linear, ordinary differential equations
\begin{eqnarray}
&&\frac{d\delta}{dr}=-r\left(\frac{d\varphi}{dr} \right)^2, \label{ElDelta} \\
&&\frac{d m}{dr}=r^2\left\{\frac{1}{2}f\left(\frac{d\varphi}{dr} \right)^2 -
   {\cal A }^{4}(\varphi) \left[L(X) -  2X\partial_{X}L(X)\right]  \right\}, \label{ElM}\\
&&\frac{d }{dr}\left( r^{2}f\frac{d \varphi}{dr} \right)=r^{2} \left\{  -
4\alpha (\varphi){\cal A}^{4}(\varphi) \left[L(X) -  X\partial_{X}L(X)\right] -r f\left(\frac{d\varphi}{dr} \right)^3\right\}.\label{ElPhi}
\end{eqnarray}

\subsection{Qualitative investigation}

Some general properties of the solutions in both the magnetically and the electrically charged cases
can be derived through an analytical investigation of the equations.
We will use the fact that for the nonlinear electrodynamics under consideration the following relation holds:
\begin{equation}
 X\partial_{X}L(X)- L(X)>0\label{EDHAM}.
\end{equation}
This inequality holds for a large class of nonlinear electrodynamics  including the
Euler-Heisenberg electrodynamics given by Lagrangian (\ref{EHL}).

The non-existence of inner horizons for our solution can be proved through an analytical analysis of
the following equation %
\begin{equation}
\frac{d }{dr}\left( e^{-\delta}r^{2}f\frac{d\varphi }{dr} \right)=4 r^2 e^{-\delta} \alpha(\varphi)
{\cal A}^{4}(\varphi)\left[X\partial_{X}L(X)- L(X)\right]>0,\label{phianl} \\
\end{equation}
which is another form of equations (\ref{MagnPhi}) and (\ref{ElPhi}).
Let us admit that more than one horizon exists.
Then, we integrate eq.(\ref{phianl})
in the interval $r\in[r_{-},r_{+}]$ where we denote he first inner and the outer horizons by $r_{-}$
and $r_{+}$, respectively, i.e.,
\begin{eqnarray}
&&\left. \left( e^{-\delta}r^{2}f\frac{d\varphi }{dr} \right) \right|_{r_{+}}-
\left.\left( e^{-\delta}r^{2}f\frac{d\varphi }{dr} \right) \right|_{r_{-}}\nonumber \\
&&\hspace{1cm}=4 \int\limits_{r_{-}}^{r_{+}}r^2 e^{-\delta} \alpha(\varphi)
{\cal A}^{4}(\varphi)\left[X\partial_{X}L(X)- L(X)\right] dr>0, \nonumber  \\ \label{varhu}
\end{eqnarray}
Since $f(r_{-})=0=f(r_{+})$ the left-hand side (LHS) of (\ref{varhu}) is equal to zero.
The integral on the RHS, however, is positive. The contradiction we reach means that our
admission is incorrect.
So if a black hole exists it will have a single horizon, i.e., its causal structure will be Schwarzschild-like.
In both conformal frames, inside the event horizon a space-like singularity is hidden.

The qualitative behavior of $\delta(r)$ can easily be seen from eqs.(\ref{MagnDelta}) and (\ref{ElDelta}).
It decreases monotonously with $r$ for both the magnetically and the electrically charged cases.

The scalar field plays a crucial role in changing the causal structure of the magnetically charged black hole.
In GR,
for $(M/P)^2\leq 24/25$ a single horizon exists, but for
$(M/P)^2>24/25$ a second and a third horizons occur. Extremal solutions exist only for
$b^2\leq b_{\rm crit}^2=8/27P^2$. In the presence of the scalar field, for $b^2>b^2_{\rm crit}$,
the causal structure is qualitatively
the same as in the General Theory. For $b^2 \leq b^2_{\rm crit}$, however, the scalar field changes
the causal structure significantly and the number of horizons reduces to one.

For the electrically charged case in GR extremal solutions exist for all values of $b^2$. 
An inner horizon emerges for
$ M <M_{0}$, where

\begin{equation}
M_{0}={\Gamma ({1\over 4}) \over 2\Gamma({3\over 2})} {Q ^{{3 \over 2}} \over (2b^2)^{{1\over4}} }. \\
\end{equation}
Again, in the electrically charged case with the presence of the scalar field the black hole has a
single horizon, namely the event horizon.

The existence of extremal solutions is
possible only for theories in which $\alpha(\varphi)$ turns to zero on the horizon. This can be easily proved
through eq.(\ref{phianl}). For the extreme solution $f(r_{H})=f'(r_{H})=0$. Thus, the LHS of
(\ref{phianl}) turns to zero on the horizon. Since relation (\ref{EDHAM}) holds and $0<{\cal A}(\varphi)<\infty$, the only chance
that the RHS of the solution turns to zero is $\alpha[\varphi(r_{H})]=0$.

\section{Numerical results}

The nonlinear systems (\ref{MagnDelta})-(\ref{MagnPhi}) and (\ref{ElDelta})-(\ref{ElPhi}) are
inextricably coupled and the event horizon $r_H$ is {\it a priori} unknown boundary. In order to be solved, they are
recast as a equivalent first order systems of ordinary differential equations.
Following the physical assumptions of the matter under consideration the asymptotic boundary conditions are set, i.e.,
$$\lim_{r \to \infty}m(r) =M \quad (M \>{\rm is\> the\> mass\> of\> the\> black\> hole\> in\> the\> Einstein\> frame}),$$
$$ \lim_{r \to \infty}\delta(r)=\lim_{r \to \infty}\varphi(r)=0.$$
At the horizon both the relationship $$f(r_H)=0$$
and
the regularization condition
$$\left.\left(\frac{df}{dr}\!\cdot\! \frac{d \varphi}{d r}\right)\right|_{r=r_H} =\left.
\left\{ 4 \alpha(\varphi) {\cal A}^4(\varphi) [X \partial_X L(X)-L(X) ]\right\}\right|_{r=r_H}$$ must be held.
Supplying the governing equations (\ref{ElDelta})-(\ref{ElPhi}) with the above five conditions we compose
well-posed boundary-value problem (BVP) for functions $\delta(r)$, $m(r)$, $\varphi(r)$ as well as the spectral
parameter $r_H$. We treat it by the Continuous Analog of
Newton Method (see, for example \cite{gavurin},\cite{jidkov},\cite{YFBT}). After an appropriate
linearization
we render the original BVP to solving a vector two-point BVP. On a discrete level sparse (almost diagonal) linear algebraic
systems with regard to increments of sought functions $\delta(r)$, $m(r)$, and $\varphi(r)$ have to be inverted.

For our numerical solution we have considered theories with constant coupling parameter, which correspond to
the Brans-Dicke theory. So the coupling function has the following form:
\begin{equation}
{\cal A}(\varphi)=e^{\alpha\varphi}, \\
\end{equation}
where $\alpha$ is a positive constant and in this theory $\alpha(\varphi)=\alpha={\rm const}$ .
We have studied the parametric space for fixed value of the coupling parameter $\alpha = 0.01$
(this value is close to the one established on the bases of experimental data) and for
several values of the magnetic charge.

\subsection{Magnetically charged case}

\begin{figure}[htbp]%
\vbox{ \hfil \scalebox{1.0}{ {\includegraphics{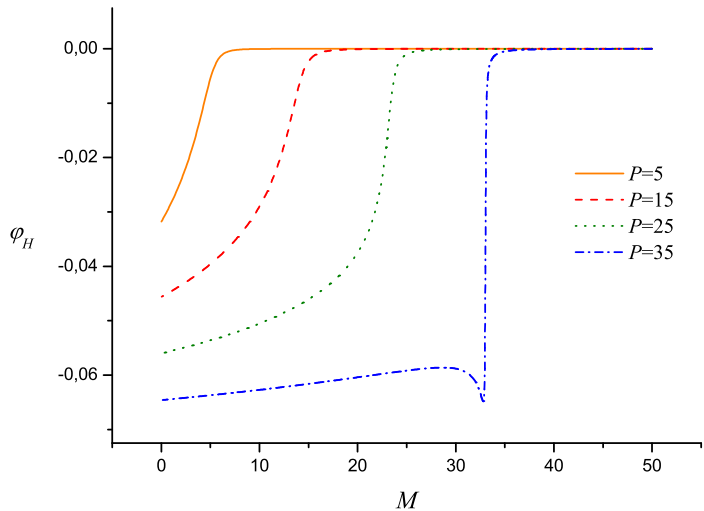}} }\hfil}%
\bigskip%
\caption{%
The value of scalar field $\varphi$ on the horizon as a function of the mass $M$, for
$P=5, 15, 25, 35$ in the magnetically charged case. For values of $(P/M)^2$ greater
than the critical $24/25$ the absolute value of the
scalar field increases considerably and prevents the formation of a degenerate horizon.} \label{PhiM}%
\end{figure}%

\begin{figure}[htbp]%
\vbox{ \hfil \scalebox{1.0}{ {\includegraphics{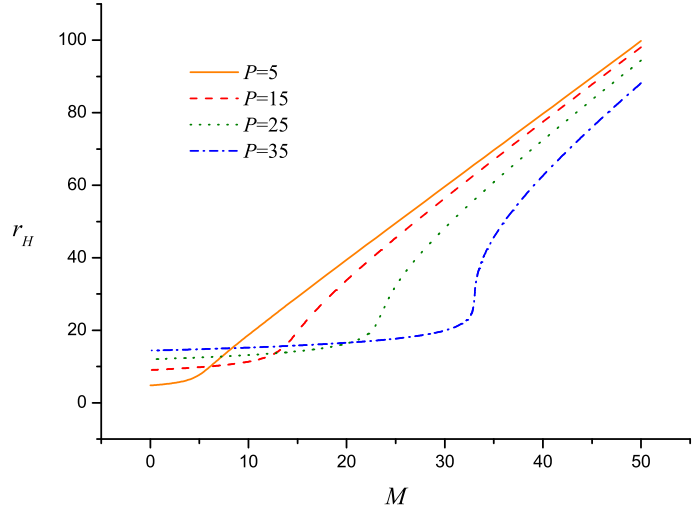}} }\hfil}%
\bigskip%
\caption{%
The $ M-r_{H} $ relation in the magnetically charged case, for the same value of the parameters
as in Fig.(\ref{PhiM}).
As the mass approaches zero, the radius of the black hole approaches
a finite value and becomes almost constant with mass $M$.
} \label{RhM}%
\end{figure}%

 \begin{figure}[htbp]%
\vbox{ \hfil \scalebox{1.0}{ {\includegraphics{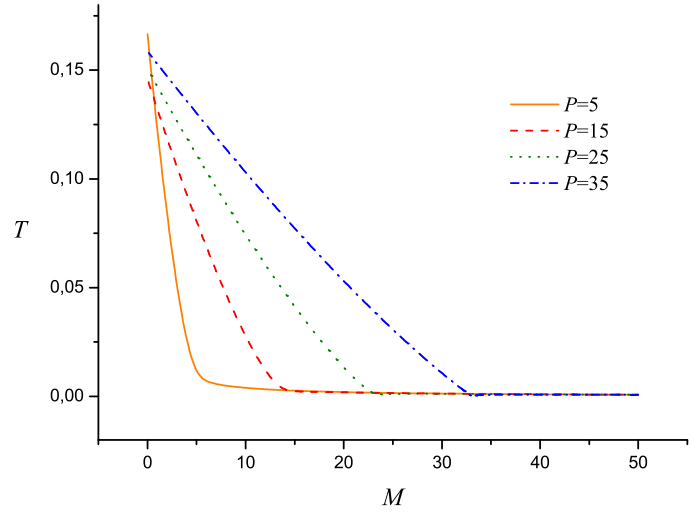}} }\hfil}%
\bigskip%
\caption{%
$ M-T $ relation in the magnetically charged case for the same value of the parameters
as in Fig.(\ref{PhiM}). A magnification of the flat region of
the curves is shown in Fig.(\ref{UvelT}). } \label{TMagn}%
\end{figure}%

 \begin{figure}[htbp]%
\vbox{ \hfil \scalebox{1.0}{ {\includegraphics{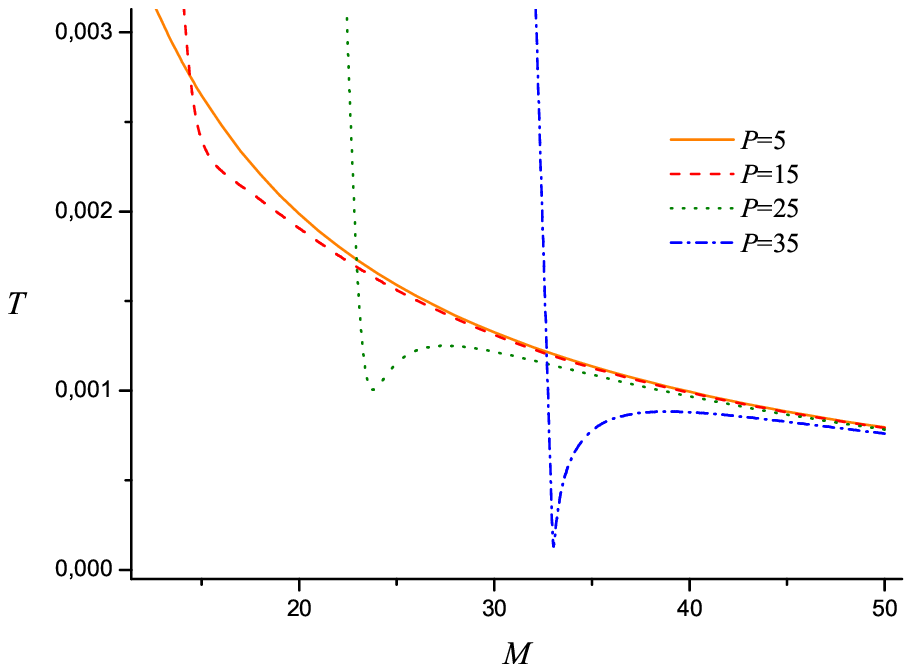}} }\hfil}%
\bigskip%
\caption{%
A magnification of the flat region of the curves in Fig.(\ref{TMagn}) . %
} \label{UvelT}%
\end{figure}%

For regions where the ratio $(P/M)^2$ is considerably less than the critical value $ 24/25 $ the behavior of the solutions
resembles that the GR case. When we decrease the mass for a fixed value of $P$, however,
the solutions starts to deviate from
it considerably. In GR the solutions pass through an extremal black hole \cite{YT}. In the case we consider,
the absolute value of the scalar field rises considerably and prevents the emergence
of a degenerate horizon (see Fig.(\ref{PhiM})). 

The numerical investigations also reveal that as the mass approaches zero, the radius of the black hole approaches
a constant value, which means that solutions with negative masses and finite radius of the event horizon exist Fig.(\ref{RhM}).

The temperature of the magnetically charged black hole is shown in Fig.(\ref{TMagn}).
As it can be seen in Fig.(\ref{UvelT}), which is a magnification of the flat region of
the curves in Fig.(\ref{TMagn}), when the mass  $M$ approaches the critical value, the temperature
decreases but rises suddenly before it reaches the zero. So an extremal solution is not reached.

\subsection{Electrically charged case}

\begin{figure}[htbp]%
\vbox{ \hfil \scalebox{1.0}{ {\includegraphics{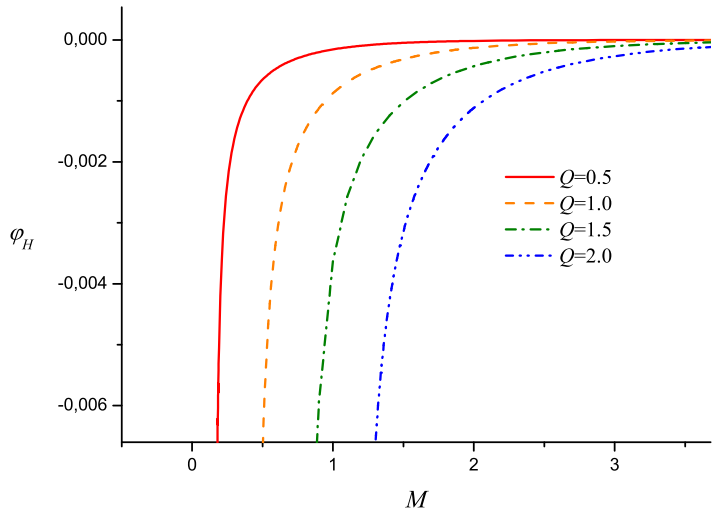}} }\hfil}%
\bigskip%
\caption{%
The value of scalar field $\varphi$ on the horizon as a function of the mass $M$, for
$Q=1.6, 2.0, 3.0$ in the electrically charged case.%
} \label{PhiEl}%
\end{figure}%

 \begin{figure}[htbp]%
\vbox{ \hfil \scalebox{1.0}{ {\includegraphics{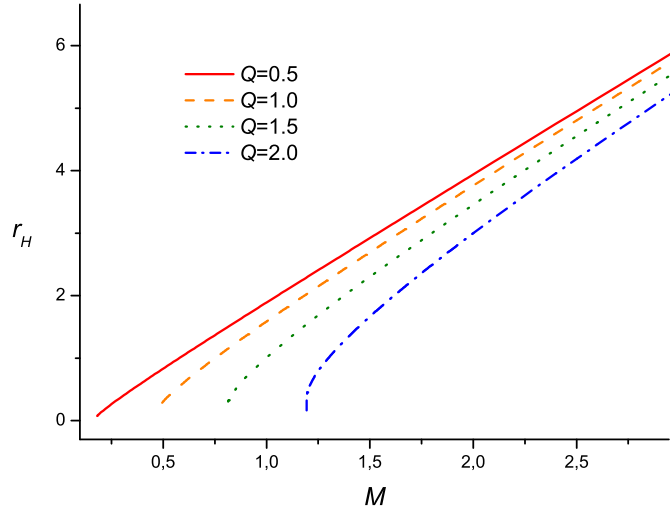}} }\hfil}%
\bigskip%
\caption{%
The $ M-r_{H} $ relation in the electrically charged case, for the same value of the parameters
as in Fig.(\ref{PhiEl}). Unlike the magnetically charged case, the radius of the horizon turns to zero
for a finite value of the mass of the black hole.%
} \label{RhEl}%
\end{figure}%

 \begin{figure}[htbp]%
\vbox{ \hfil \scalebox{1.0}{ {\includegraphics{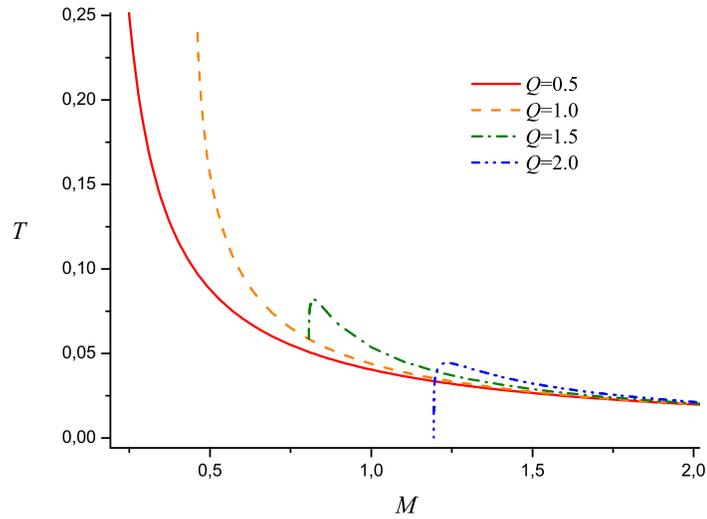}} }\hfil}%
\bigskip%
\caption{%
In this figure, the $ M-T $ relation for the same value of the parameters
as in Fig.(\ref{PhiEl}) is shown. %
} \label{TEl}%
\end{figure}%

As in the magnetically charged black hole in the electrically charged one the causal structure is simpler than in the
corresponding solution in GR. Instead of having an inner and outer horizon,
which merge in a single degenerate horizon for some value of the mass, as in the GR case,
the electrically charged black hole has a
single horizon.
The properties of the electrically charged black hole, however, differ significantly from those of the magnetically
charged case.
The radius of the electrically charged black hole decreases with the decrease of its mass, and solutions with
negative masses
do not exist (see Fig.(\ref{RhEl})). The behavior of the temperature,
however, is complex as this can be seen in Fig.(\ref{TEl}).  Due to numerical difficulties
which arise in this region of the parametric space, some of the curves are incomplete.
According to the curves we have obtained, for some values of the electrical charge $Q$ the temperature
approaches zero. A possibility exists, that with the decrease of the mass the temperature passes through a local minimum
and then increases to infinity similarly to the magnetically charged case. To answer this question we should carry out
some analytical investigations. In cases of low mass the radius of the black hole is small which allows us to make an
approximation in the equations by leaving only the leading terms on the RHS. From the approximate system
the following relation can be derived
\begin{equation}
r^2e^{\delta} \left[  \alpha \left(e^{-2\delta}f \right)'-  \left(e^{-2\delta}f \right) \varphi'\right]=C,
\end{equation}
where $C$ is an arbitrary positive constant.
On the horizon this relation takes the following form
\begin{equation}
\alpha r_{H}^2 e^{-\delta_{H}}f'_{H}=C.
\end{equation}
Hence, for the temperature of the electrically charged black hole we obtain
\begin{equation}
T=\frac{1}{ 4\pi C \alpha r_{H}^2 },\label{Temp}
\end{equation}
from where we can say that it rises to infinity as $r_{H}$ and $M$ approach zero, for all values of the parameters.

\section{Thermodynamics}

For the solution we study the First Law (FL) of thermodynamics holds.
It is naturally connected with the Einstein frame as this can be seen below.
The formal derivation of the FL of black
hole thermodynamics for the case of non-linear electrodynamics
including the presence of a scalar field, in the Einstein frame, can be seen in the work of Rasheed \cite{Rasheed}.
The formulation of the FL in the
Jordan frame requires proper definition of the thermodynamic variables.

The temperature of the event horizon is invariant under
conformal transformations of the metric that are unity at infinity \cite{Jacobson}.

The properly defined entropy is also preserved under conformal transformations.
In the Einstein frame, the entropy of the black hole is one forth of
the horizon area. In the Jordan frame, however,
this is not so \cite{MVisser,FordRoman} and the definition of the entropy needs to be generalized as follows:
\begin{equation}
S_{J}=\frac{1}{4G_{*}}\int d^2x \sqrt{-^{(2)}{\tilde g}}F(\Phi).
\end{equation}
Using relation (\ref{conf}) we find that
\begin{equation}
S_{J}=\frac{1}{4G_{*}}\int d^2x \sqrt{-^{(2)} g}=S_{E}=S.
\end{equation}
In the last two equations quantities $^{(2)}{\tilde g}$ and $^{(2)} g$ are the determinants of the induced metrics on the horizon
in the Jordan
and in the Einstein frame, respectively.

The presence of the scalar field adds a term ${\cal{D}}\delta\varphi_{\infty}$
in the FL, where ${\cal{D}}$ is the scalar field charge and $\delta\varphi_{\infty}$ is the variation of the
asymptotic value of the scalar field.
We define the scalar field charge as
\begin{equation}
{\cal{D}} = - \left. r^2 \frac{d\varphi}{dr}\right|_{r\rightarrow \infty}. \label{dilatoncharge}
\end{equation}
This charge, however,
is not independent. Let us integrate
eq.(\ref{phianl}) from the radius of the horizon to infinity. We obtain
\begin{equation}
{\cal{D}}=4 \int\limits_{r_{H}}^{\infty} r^2 \alpha(\varphi)
{\cal A}^{4}(\varphi)\left[X\partial_{X}L(X)- L(X)\right]dr.
\end{equation}
Hence, we see that the scalar field charge can be determined unambiguously once the mass and the magnetic charge of
the black hole and the
asymptotic value of the scalar field at infinity are known.
In our case, since the scalar field is fixed and vanishing at infinity the term coming from the
scalar field disappears.

Scalar-tensor theories of gravity violate the strong equivalence principle. This
results in the appearance of three different possible masses as a measure of the
total energy of the compact objects.
In both conformal frames in the FL of thermodynamics the Arnowitt-Deser-Misner (ADM) mass
from the Einstein frame should be used. Similarly, for boson and fermion stars the proper measure for the
energy of the system is again the
ADM mass in the Einstein frame $M$. For more details on the subject
we would refer the reader to the works \cite{Lee,Shapiro,Whinnett,Yazadji}.

The ADM masses in both frames are not equivalent. The
ADM mass in the Jordan frame $M_{J}$ is related to the
ADM mass in the Einstein frame $M$, and the scalar field charge ${\cal D}$ in the following way

\begin{equation}
M_{J}=M+\alpha\cal{D}.
\end{equation}
It can be interpreted as the Keplerian mass of the black hole.

The terms in FL connected with the magnetic and the electric charges, respectively also remain invariant
with the transition between the two conformal frames.

To sum up, the properly defined FL looks in the same way in both conformal frames. Its explicit
form for the magnetically charged case and for the electrically charged case, respectively, is presented below.

\subsection{Magnetically charged case}

In the magnetically charged case the FL of thermodynamics is
\begin{equation}
\delta M=T \delta S + \Psi_{H}\delta P,\label{FirstLawM}
\end{equation}
where $T$, $S$ and $P$ are the temperature, the entropy, and the magnetic charge of the black hole. The
quantity $\Psi$ conjugate to the magnetic charge  is the potential of the magnetic field which is given by the
following definition
\begin{equation}
H_{\mu}=\partial_{\mu}\Psi.\label{defH}
\end{equation}
On the other hand the magnetic field is defined as
\begin{equation}
H_{\mu}=-\star G_{\mu\nu}\xi^{\nu},
\end{equation}
where
\begin{equation}
G_{\mu\nu}=-\frac{1}{2} \frac{\partial\left( {\cal A}^{4}(\varphi)L\right)}{\partial F_{\mu\nu} },
\end{equation}
$\xi=\frac{\partial}{\partial t}$ is the Killing vector generating time translations
and ``$\star$'' is the Hodge star operator.

\subsection{Electrically charged case}

In the electrically charged case the FL of black hole thermodynamics takes the following form
\begin{equation}
\delta M=T \delta S + \Phi_{H}\delta P,\label{FirstLawEl}
\end{equation}
where $T$, $S$ and $Q$ are the temperature, the entropy, and the electric charge of the black hole. The
quantity $\Phi$ conjugate to the electric charge is the potential of the magnetic field which is given by the
following definition
\begin{equation}
E_{\mu}=\partial_{\mu}\Phi.\label{defH}
\end{equation}
On the other hand the electric field is defined as
\begin{equation}
E_{\mu}=F_{\mu\nu}\xi^{\nu}.
\end{equation}

\section{Conclusion}

In the present work numerical solutions describing charged black holes coupled to non-linear electrodynamics in
the scalar-tensor theories with massless scalar field were found. Purely magnetically and purely electrically charged
cases were studied. For the Lagrangian of the non-linear electrodynamics the truncated Euler-Heisenberg
Lagrangian was chosen and scalar-tensor theories with positive coupling parameter were considered. As a result of the
numerical and analytical investigation, some general properties of the solutions were found. Both the
magnetically charged case and electrically charged case have a single, non-degenerate event horizon, i.e., their causal
structure resembles that of the Schwarzschild black hole and is simpler compared to the corresponding solution in the
frame of GR. In both conformal frames, inside the event horizon a space-like singularity is hidden.

Some properties of the purely magnetically charged black hole differ
significantly from those of the electrically charged one. As the mass decreases, the radius of the horizon in the
first case approaches a finite, constant value and the temperature of the horizon rises. Thus, solutions with negative
masses and finite radius of the event horizon exist. In the electrically charged case, however,
the radius of the horizon turns to zero for a finite value of the mass $ M $ of the black hole.
In this case, in the $r_{H}\rightarrow 0$ limit a naked singularity is reached and the
solution cannot be continued for negative masses.

\section*{Acknowledgments} This work was partially supported by
the Bulgarian National Science Fund under Grant MUF04/05 (MU 408)
and the Sofia University Research Fund N60.

\end{document}